# Human Resource Management System

## A.S.Syed Navaz[1,]  A.S.Syed Fiaz[2], C.Prabhadevi[3], V.Sangeetha[4], S.Gopalakrishnan[5]

[1,3,4]*Department of Computer Applications, Muthayammal College of Arts & Science/ Periyar University, India*
[2]*Department of Computer Science, Sona College of Technology/ Anna University, India.*
[5]*Department of Computer Science, Muthayammal College of Arts & Science/ Periyar University, India*

**Abstract:** *The paper titled **"HUMAN RESOURCE MANAGEMENT SYSTEM"** is basically concerned with managing the Administrator of HUMAN RESOURCE Department in a company. A **Human Resource Management System (HRMS)**, refers to the systems and processes at the intersection between human resource management (HRM) and information technology. It merges HRM as a discipline and in particular its basic HR activities and processes with the information technology field, whereas the programming of data processing systems evolved into standardized routines and packages of enterprise resource planning (ERP) software[1]. The main objective of this paper is to reduce the effort of Administrator to keep the daily events such as attendance, projects, works, appointments, etc.*

*This paper deals with the process of identifying the employees, recording their attendance hourly and calculating their effective payable hours or days. This paper should maintain the records of each and every employee and their time spend in to company, which can be used for performance appraisal. Based on that transfer, removal, promotion can be done.*

## I. Introduction

The paper is used to maintain efficiently the HR department schedule of any type of company. In larger organization, employees are large. At that time this paper is useful and helpful. HR Management system is not only becomes a desire of the company but it becomes the need of the company. The Administrator gets into the system using admin name and a password.

**1.1 Advantages**
**1.** Easy access to the data
**2.** The new system is more user-friendly, reliable and flexible.
**3**. Data alteration is easy.
**4.** Maintenance of the project is easy.
**5.** Reduced manual work.
**6.** Timely Report generation.

The main objective of this paper is to reduce the effort of administrator to keep the daily events such as payroll, employee performance, and employees' details. It consists of six modules.

They are:
1. Employee Details
2. Payroll
3. Training
4. Performance
5. Resignation
6. Resume tracking.

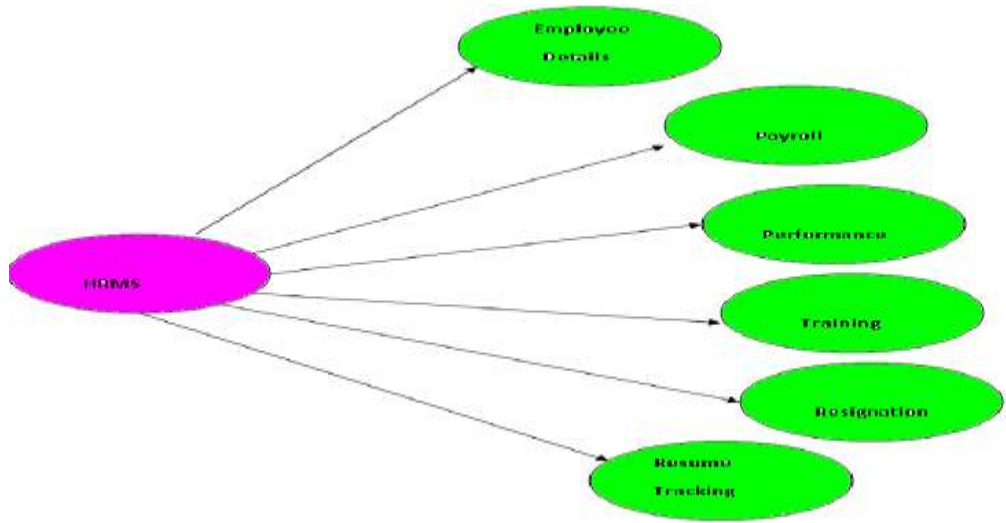

Fig 1: Overview of Human Resource Management Systems

**1.2 Employee Details**

Employee Details module is used to maintain the employees' details such as adding new employee, modifying the existing employee and deleting the existing employee. When a new employee is selected from the resume tracking, all the details are to be entered and maintained in the database.

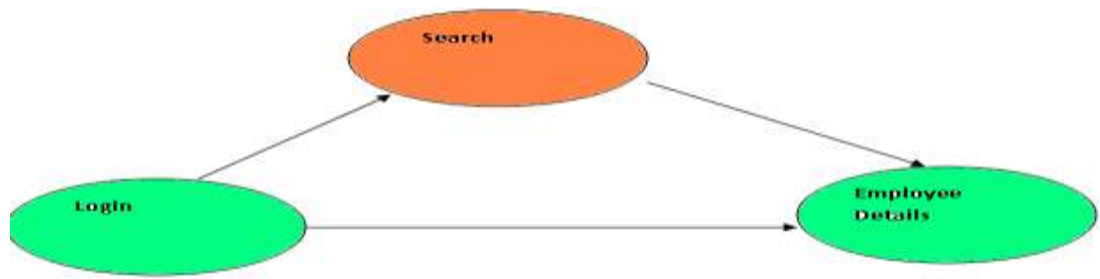

Fig 2 : Overview of Employee Details

The employee details contains three kind of information.
1. Personal Information
2. Contact Information
3. Employee Status

In the personal information, it consists of the information about the employee name, employee id, nationality, etc. In the contact information, it consists of the information about the employee address, phone numbers, etc,

In the employee status, it consists of the information about the status of the employee, supervisor name, department, etc.

**1.3 PAYROLL**

In the payroll module, it consists of the information about the employee salary details such as basic pay, allowances, deductions and calculate the gross pay and net pay from the given allowances and deductions.

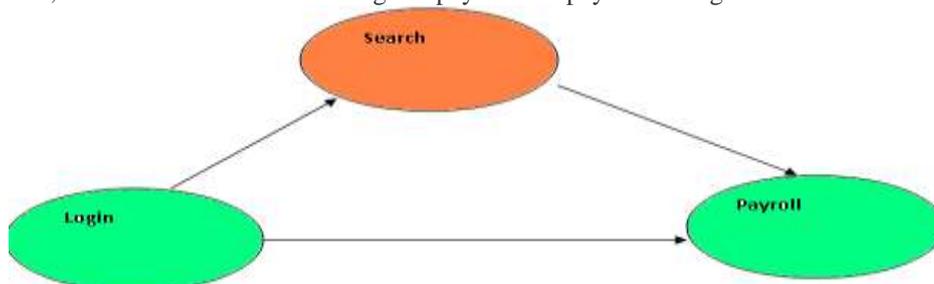

Fig 3 : Overview of Payroll Details

All the employees' pay details are maintained by the HR manager. The main function of this payroll module is to maintain the employee pay information.

## 1.4 TRAINING

In this training module, it consists of the employees' schedule about the training conducted in the organization for the particular employee. The employees' previous training experience will be maintained in the database.

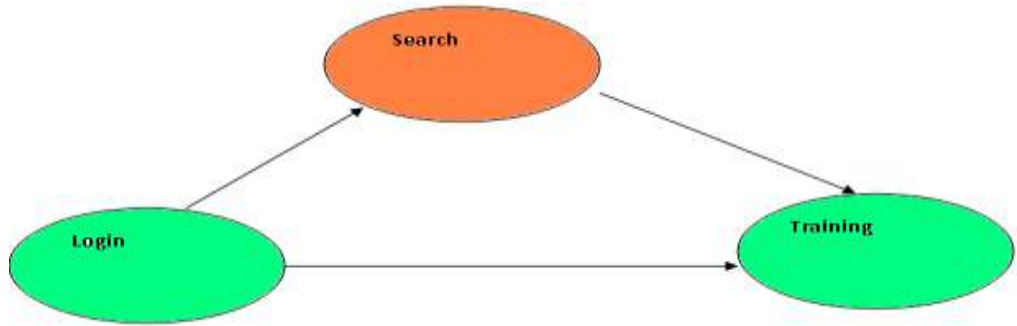

Fig 4 : Overview of Training Module

In the module contains the information about the employee who are in the Training and who are finished the training. These details are to be used in the payroll calculation.

## 1.5 PERFORMANCE

This performance module contains information about the employee's current position in the organization. This module has the information such as employee name, employee ID, Division, work group, evaluation date, evaluator, and evaluation period.

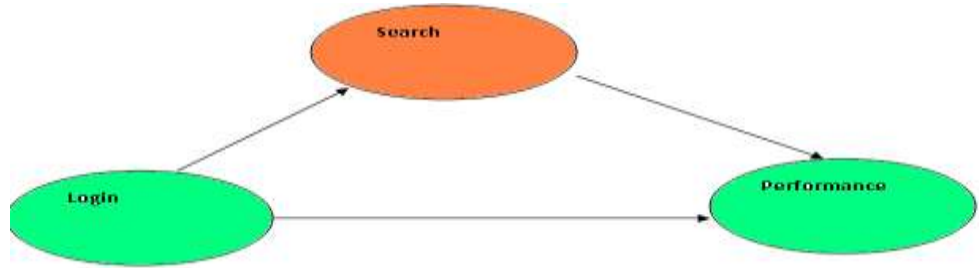

Fig 5 : Performance of Employees Poisition

This module is used to monitor the employees, their work performance and the involvement of them in the organization.

### 1.5.1 Leave Management

This module contains the information about the employees leave details. There are three kinds of leave which are sick leave, vacation, and holiday.

There are fixed amount of days that are allocated for each type of leave and the database of leave details are maintained by the organization. The details includes number of days, period, total number of leave taken by that employee upto that date and number of days that are remaining.

## 1.6 Resignation

This module contains the information about the ex-employees who have worked for the organization. The information are such as department, position, their supervisor, current contact information, joining date and resigning date.

These information are used to contact the ex-employees in case of emergency in which project they have already involved.

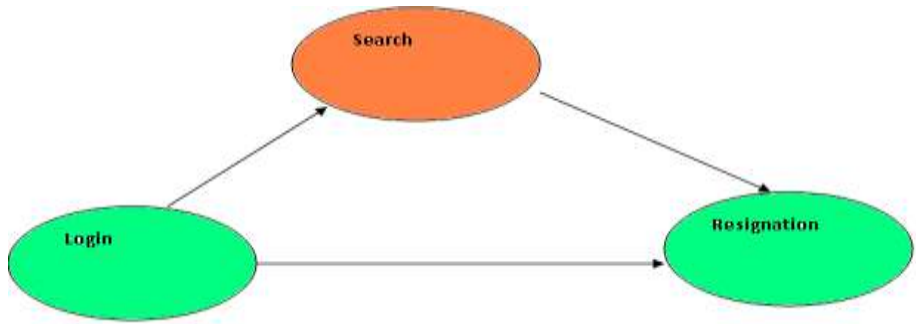

Fig 6 : Ex-Employee Details

**1.7 RESUME TRACKING**
This module contains the information about the applicants such as their Curriculum, their contact information, their work experience, area of specialization and area of interest.

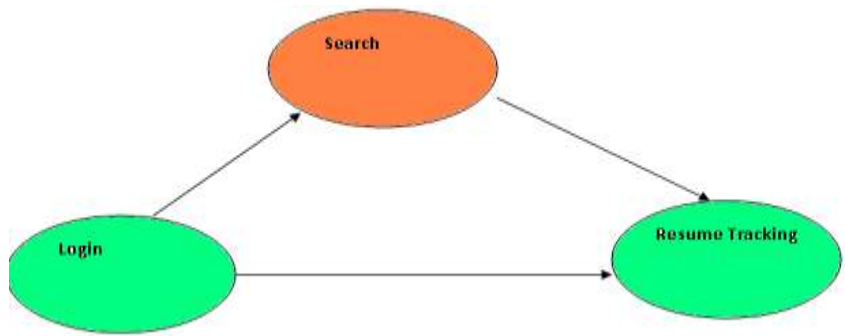

Fig 7 : Resume Tracking

The applicant have the facility of registering their resume through online and their details are stored in the organizations database. If the applicants details suits the organizations requirement then they can be called for next step.

## II. Tables

**2.1 LOGIN**

| ATTRIBUTE | DATATYPE | CONSTRAINTS | DESCRIPTION |
|-----------|----------|-------------|-------------|
| Userid | Varchar2(10) | Primary key | User Id |
| password | Varchar2(20) | Not Null | Password |

## 2.2 EMPLOYEE

| ATTRIBUTE | DATA TYPE | CONSTRAINTS | DESCRIPTION |
|-----------|-----------|-------------|-------------|
| Title | Varchar2(20) | Not Null | Title |
| Empid | Varchar2(10) | Primary key | Employee ID |
| Firname | Varchar2(10) | Foreign Key | First Name |
| Midname | Varchar2(20) | Null | Middle Name |
| Lastname | Varchar2(20) | Not Null | Last name |
| Blood | Varchar2(20) | Not Null | Blood group |
| Nation | Varchar2(20) | Not Null | Nationality |
| Address | Varchar2(20) | Not Null | Address |
| City | Varchar2(20) | Not Null | City |
| State | Varchar2(20) | Not Null | State |
| Pin | Number | Not Null | Pin code |
| Home | Number | Not Null | Home phone |
| Workplace | Number | Not Null | Work Place phone |
| Mobile | Number | Null | Mobile Number |
| Email | Varchar2(20) | Not Null | Email ID |
| Status | Varchar2(20) | Not Null | Status |
| Supervisor | Varchar2(20) | Not Null | Supervisor Name |
| Hdate | Date | Not Null | Hire Date |
| Dept | Varchar2(20) | Not Null | Department |
| Bdate | Date | Not Null | Birth Date |
| gender | Char | Not Null | Gender |
| marital | Char | Not Null | Marital Status |

## 2.3 LEAVE MANAGEMENT

| ATTRIBUTE | DATA TYPE | CONSTRAINTS | DESCRIPTION |
|-----------|-----------|-------------|-------------|
| empname | Varchar2(10) | Foreign key | Employee Name |
| Empid | Varchar2(10) | Primary key | Employee ID |
| vacstart | Number | Not Null | Vacation Balance Start |
| vacbalance | Number | Not Null | Vacation Balance |
| V1date | Date | Not Null | Vacation Last Taken |
| sickstart | Number | Not Null | Sick Balance Start |
| S1date | Date | Not Null | Sick last Taken |
| holstart | Number | Not Null | Holiday Balance Start |
| Holbal | Number | Not Null | Holiday balance |
| H1date | Date | Not Null | Holiday last Taken |

## 2.4PERFORMANCE

| ATTRIBUTE | DATA TYPE | CONSTRAINTS | DESCRIPTION |
|-----------|-----------|-------------|-------------|
| Empname | Varchar2(10) | Foreign key | Employee Name |
| Empid | Varchar2(10) | Primary Key | Employee ID |
| Dept | Varchar2(20) | Not Null | Department |
| Workgroup | Varchar2(20) | Not Null | Work Group |
| Division | Varchar2(20) | Not Null | Division |
| Position | Varchar2(20) | Not Null | Position |
| Evaluate | Date | Not Null | Evaluation Date |
| Evaluator | Varchar2(20) | Not Null | Evaluator |
| Revfr | Varchar2(20) | Not Null | Review Period From |
| Revto | Varchar2(20) | Not Null | Review Period To |
| responsibility | Varchar2(20) | Not Null | Responsibility |

## 2.5 RESIGNATION

| ATTRIBUTE | DATA TYPE | CONSTRAINTS | DESCRIPTION |
|-----------|-----------|-------------|-------------|
| Title | Varchar2(10) | Not Null | Title |
| Empname | Varchar2(10) | Not Null | Employee Name |
| Empid | Varchar2(20) | Primary key | Employee ID |
| position | Varchar2(20) | Not Null | Position |
| Dept | Varchar2(20) | Not Null | Department |
| Superv | Varchar2(20) | Not Null | Supervisor |
| Jdate | Date | Not Null | Joining Date |
| Rdate | Date | Not Null | Resignation Date |
| Email | Varchar2(20) | Not Null | Email ID |
| Gender | Char | Not Null | Gender |
| City | Varchar2(20) | Not Null | City |
| Homephone | Varchar2(20) | Not Null | Home Phone |

## III. Input Key

The goal of input key is to input data as accurately as possible. Here inputs are designed effectively so that the error made by operation is minimized. The input to the system has been designed and coordination in such way that there format is similar in all forms. Forms are designed in such way that relevant information is grouped together and they are placed on a single frame, so as to access easily. At the time of data entry the verification and validation of the data were done.

Input key is the most part of the overall system design, which requires very careful attention. Often the collection of the input data is most expensive part of the system. Many errors may occur during the phase of the design. So to make the system study, the inputs given by the user is strictly validated before making a manipulation with it.

## IV. Output Key

The output key is another very important phase. The outputs are mainly used to communicate with the user, processing the input data given by the user. It is documented in each stage of the project to ensure free output. The output screens are designed in very simple and easy to understand format. The quality, urgency and the frequency of outputs should be taken into consideration. All user option is presented in well-formatted forms. The quality refers to the way by which the output is presented to the user.

The reports can be used for day-to-day functioning of the business as well as management information. The reports, if generated with the specific report criteria and in a timely manner, help in operational efficiency, detecting and minimizing of errors as well as provide the pointers towards control weakness.

Output screen are designed in such a way that no ambiguity arises. Output data is presented in a well-formatted way. The required information is printed on the report in the specified format.

## V. Figures

## 5.1 LOGIN PAGE

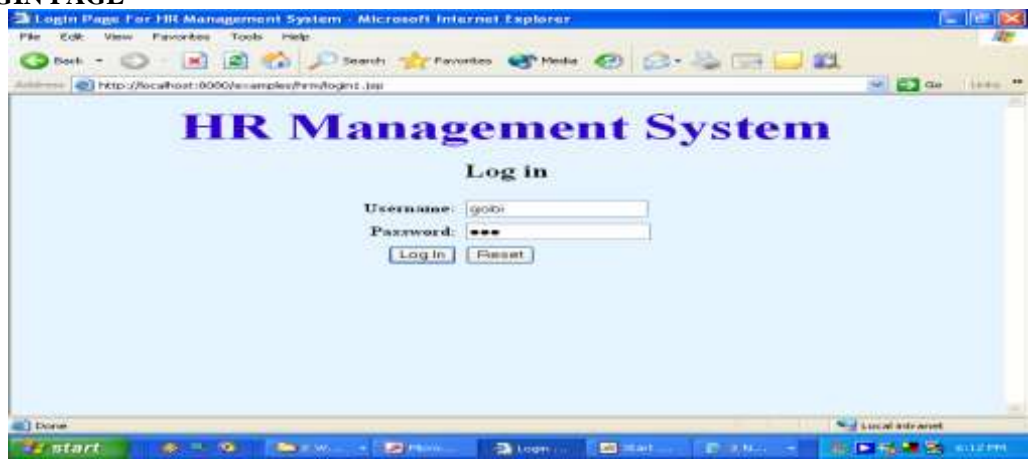

### 5.2 WELCOME PAGE

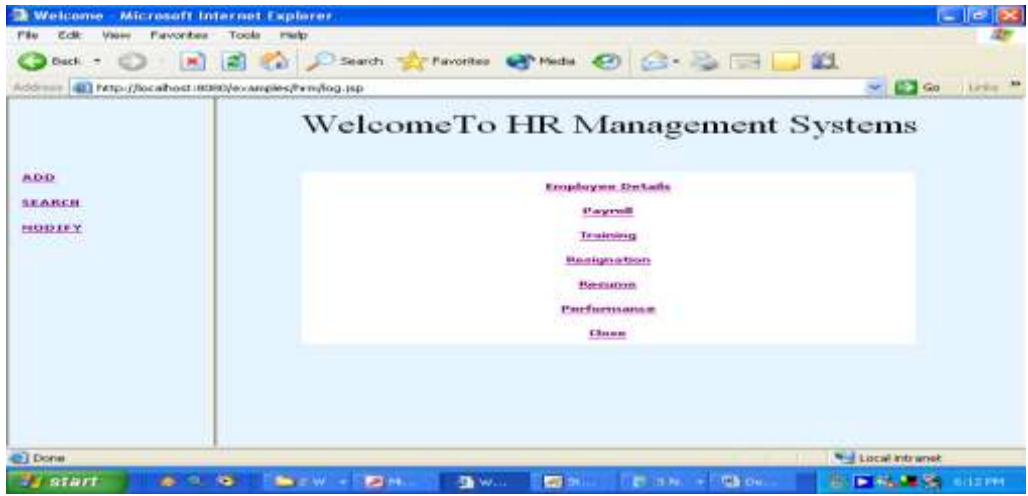

### 5.3 EMPLOYEE DETAILS

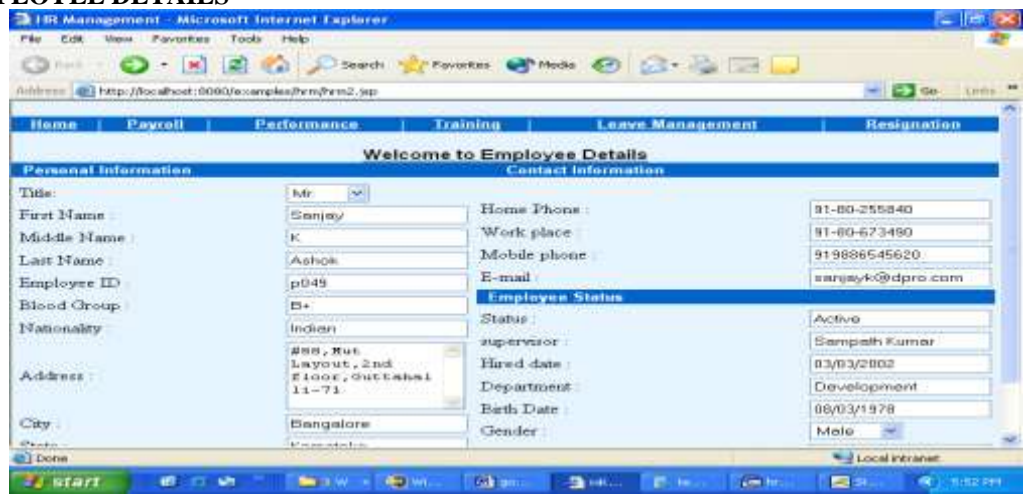

### 5.4 REGISTERED EMPLOYEE DETAILS

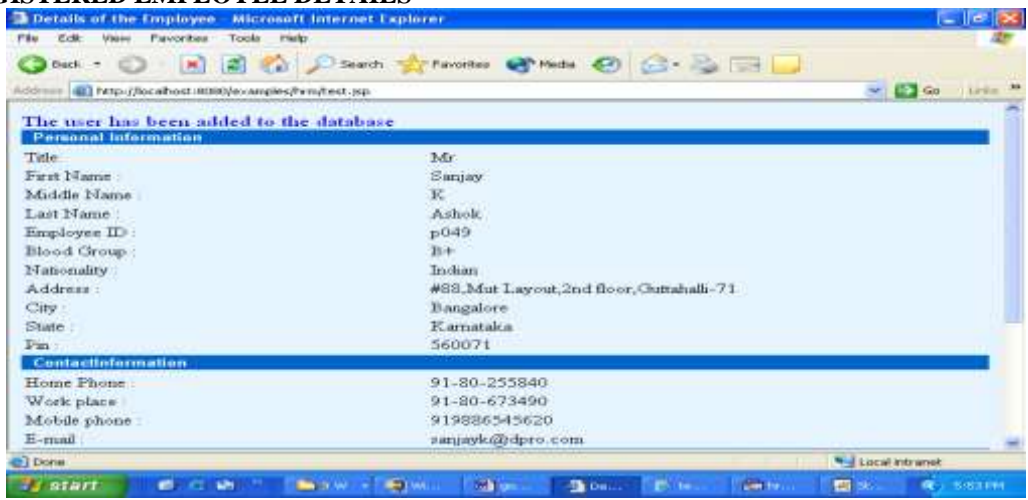

**5.5 PERFORMANCE DETAILS**

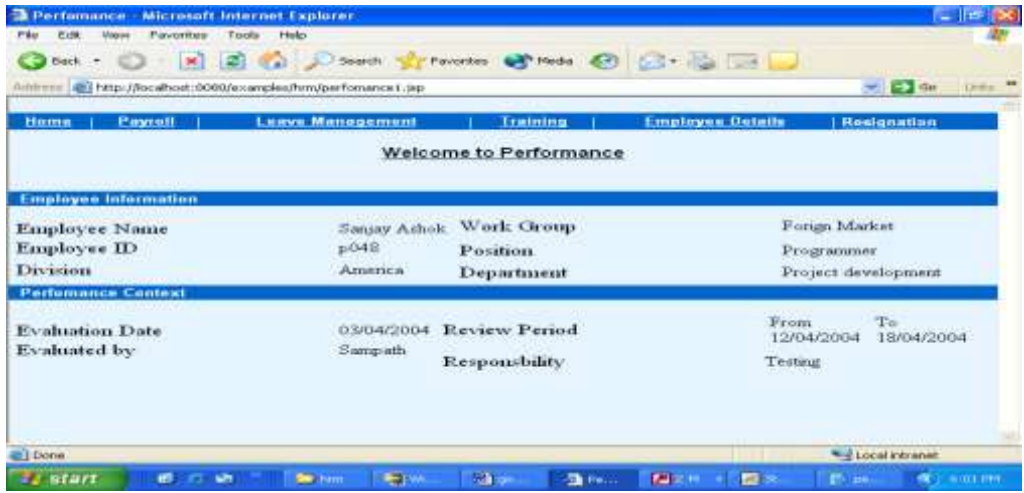

**5.6 LOGIN FOR LEAVE MANAGEMENT**

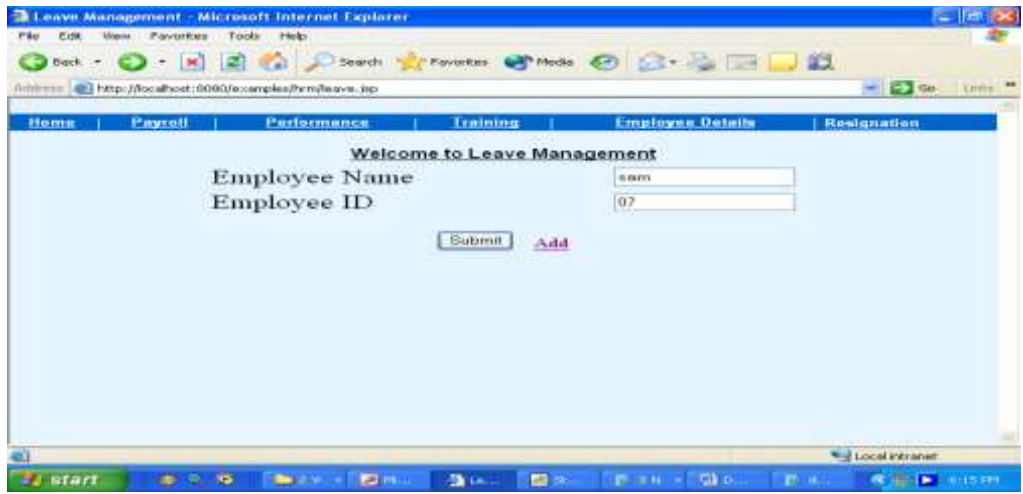

**5.7 WELCOME PAGE FOR LEAVE MANAGEMENT**

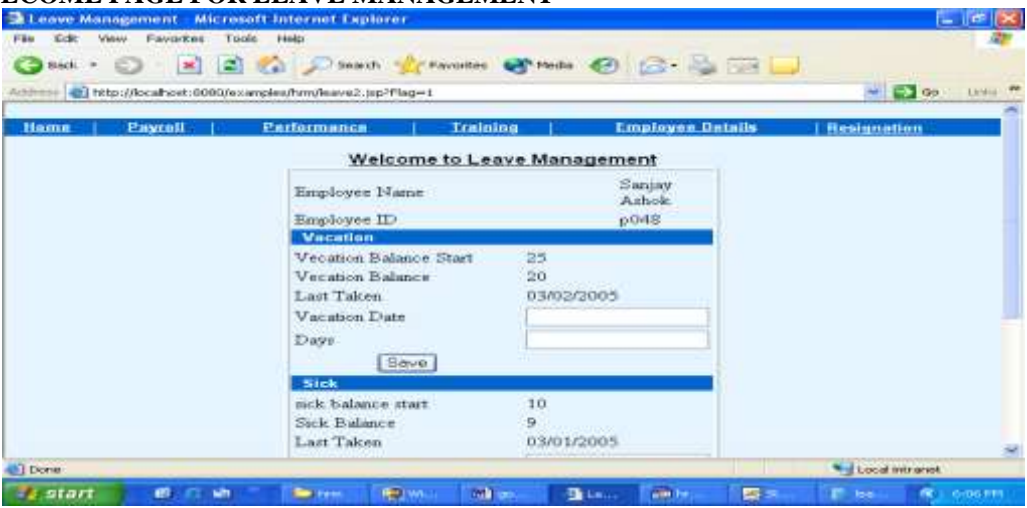

## 5.8 EX-EMPLOYEE DETAILS LOGIN PAGE

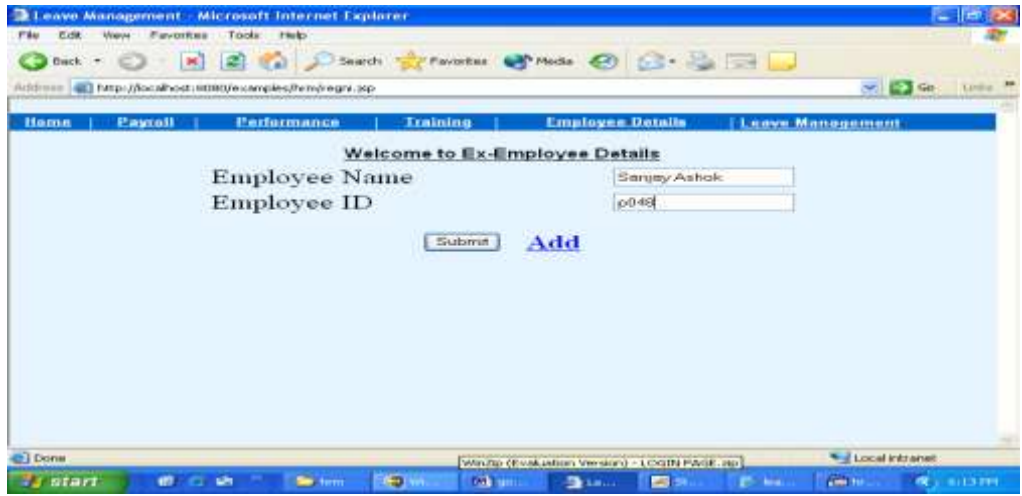

## 5.9 EX- EMPLOYEE DETAILS

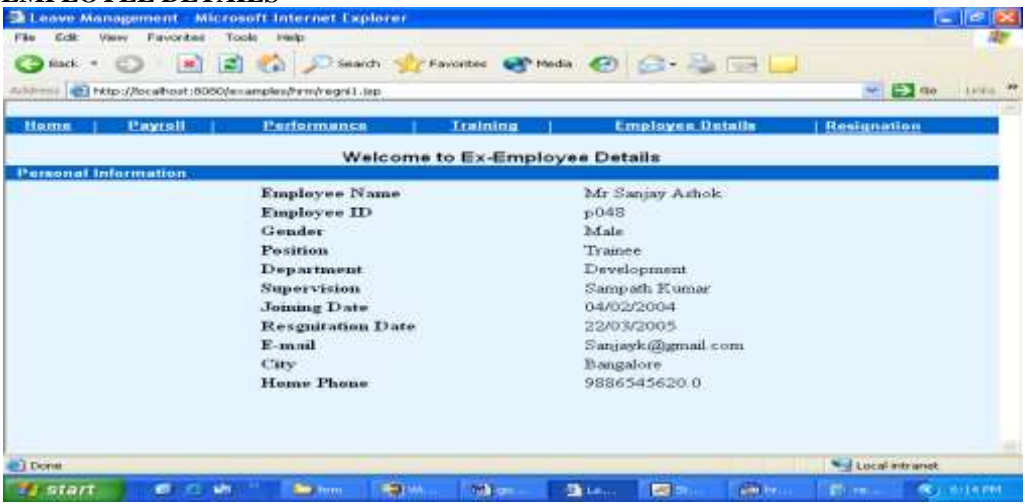

## 5.10 REGISTERATION PAGE

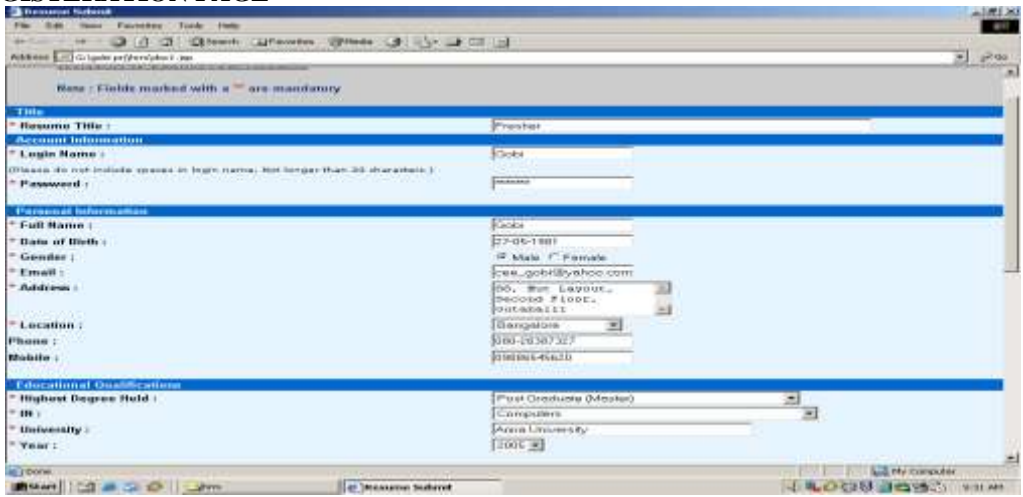

## VI. Conclusion

In conclusion I would like to tell that this **Human Resource Management Systems** has achieved its purpose. It has taken a huge task for this project to be completed. It has given a huge lift to the company's operations. What ever that has done manually has been completely shifted to the computerized process and this has enabled the company to carry out its operation more quickly. This has also given a wider spectrum of communication to the users. Since whatever that has so far been done manually has been changed to a computerized. It has resulted in more efficient processing of data.

The new system has resulted in giving numeric advantages to the company in many ways. Some of them are given below State of negligible paper work is almost reduced. Accessing and getting data can be done at a single click. Data manipulation has become simpler and the cost factor has been reduced. It is faster and more efficient processing of data. It is less time consuming. Operations are more transparency. Communications between the users is more efficient.

### Author's Information

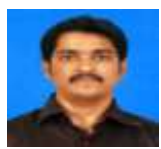

**A.S.Syed Navaz** received BBA from Annamalai University, Chidambaram 2006, M.Sc Information Technology from KSR College of Technology, Anna University Coimbatore  2009,  M.Phil in Computer Science from Prist University, Thanjavur 2010 and M.C.A from Periyar University, Salem 2010 .Currently he is working as an Asst.Professor in Department of Computer Applications, Muthayammal College of Arts & Science, Namakkal. His area of interests are Computer  Networks and Mobile Communications.

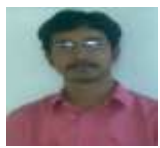

**A.S.Syed Fiaz** received BE Computer Science from Sona College of Technology, Anna University, Chennai 2010 & Currently he is pursuing his final year ME in  Computer Science from Sona College of Technology, Anna University, Chennai 2013 . His area of interests are Computer  Networks and Mobile Communications.

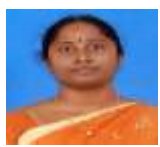

**C.Prabhadevi** received BCA from Sengunthar Arts & Science College, Periyar University 2005 , M.Sc Computer Science from Sengunthar Arts & Science College, Periyar University 2007, M.C.A from Periyar University, Salem 2009 and M.Phil in Computer Science from Prist University, Thanjavur 2012. Currently she is working as an Asst.Professor in Department of Computer Applications, Muthayammal College of Arts & Science, Namakkal. Her area of interests are Digital Image Processing and Mobile Communications.

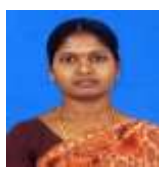

**V.Sangeetha** received B.Sc Computer Science from Mahendra Arts & Science College, Periyar University 2003, M.Sc Computer Science from Mahendra Arts & Science College, Periyar University 2005 and M.Phil in Computer Science from Periyar  University, Salem  2008. Currently she is working as an Asst.Professor  in Department of Computer Applications, Muthayammal College of Arts & Science, Namakkal. Her area of interests are Computer  Networks and Mobile Communications.

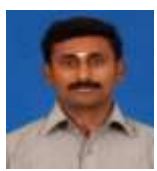

**S.Gopalakrishnan** received B.Sc Computer Science & M.Sc Computer Science from Nehru Memorial College, Bharathidasan University, Trichy 2002 & 2004, M.Phil in Computer Science from Periyar University, Salem  2009 & B.ed Computer Science from Rainbow College of Education, Namakkal 2011. Currently he is working as an Asst.Professor  in Department of Computer Science, Muthayammal College of Arts & Science, Namakkal. His area of interests are Computer  Networks and Mobile Communications.